\documentclass[prd,aps,twocolumn,showpacs,preprintnumbers,amsmath,amssymb]{revtex4}
\usepackage[dvips]{graphicx} 
\usepackage{epsf}
\usepackage{amsmath} 
\usepackage{amssymb} 
 
\usepackage{graphicx}
\usepackage{dcolumn}
\usepackage{bm}
\def\be{\begin{equation}} 
\def\ee{\end{equation}} 
\def\bea{\begin{eqnarray}} 
\def\eea{\end{eqnarray}} 

\newcommand{\comment}[1]{}

\newcommand{\tphi}{\tilde{\phi}}

\begin{document} 
 
 
\date{\today} 
 
\title{Processing of Cosmological Perturbations in a Cyclic Cosmology} 
 
\author{Robert Brandenberger$^{1,2}$} 
 
\affiliation{1) Department of Physics, McGill University, 
Montr\'eal, QC, H3A 2T8, Canada} 

\affiliation{2) Theory Division, CERN, CH-1211 Geneva, Switzerland}

 
\begin{abstract} 

The evolution of the spectrum of cosmological fluctuations from one cycle to the next
is studied. It is pointed out that each cycle leads to a reddening of the spectrum. This
opens up new ways to generate a scale-invariant spectrum of curvature
perturbations. The large increase in the amplitude of the fluctuations quickly leads
to a breakdown of the linear theory. More generaly, we see that,
after including linearized cosmological perturbations, a cyclic universe cannot
be truly cyclic.

\end{abstract} 
 
\maketitle

\newcommand{\eq}[2]{\begin{equation}\label{#1}{#2}\end{equation}} 
 
\section{Introduction} 

Recently, there has been renewed interest in cyclic cosmologies.
The motivation comes in part from general efforts to construct non-singular
cosmological backgrounds (see e.g. \cite{Novello} for a recent review), 
in part from attempts to construct
a cyclic cosmology \cite{Cyclic} as an extension of the
Ekpyrotic universe scenario \cite{Ekp}.

A problem which faces most attempts at constructing a cyclic
cosmology and which was already pointed out by
Tolman \cite{Tolman} is that the background cosmology
cannot be cyclic if the fact that entropy is generated
in each cycle is taken into account (see, however, \cite{Biswas3}
for a recent model which partially resolves this problem).
However, as discussed in \cite{BMP}, the entropy of
cosmological perturbations does not grow as long as the
fluctuations remain well described by linear theory. This is
due to the fact that each perturbation mode continues to
describe a pure state if it starts out describing a pure
state.
 
In this Note we discuss a problem for cyclic cosmology
which arises even in the absence of entropy generation:
since on super-Hubble scales the curvature fluctuations do not
evolve symmetrically - they grow in
the contracting phase and are constant in the expanding
phase - there is a net growth of the fluctuations from period
to period which destroys the cyclic nature of the
cosmology \footnote{The net growth of the amplitude of
the perturbations will eventually cause the fluctuations to
become non-linear. At that stage, we expect violent mode
mixing effects which will lead to an increase in the coarse-grained
entropy and thus to the old entropy problem for cyclic
cosmologies.}. 

Since long wavelength modes have a wavelength
which in the contracting phase is larger than the Hubble radius 
for a longer time than short wavelength modes, the spectrum
of fluctuations on super-Hubble scales in the
contracting phase is redder than the
initial spectrum on sub-Hubble scales. Thus, both the
amplitude and the slope of the spectrum of fluctuations
changes from cycle to cycle - an effect which we call
``processing of the spectrum of cosmological fluctuations".
The reddening of the spectrum takes place in each contracting
phase: the initial sub-Hubble scaling of the spectrum differs
from the later super-Hubble scaling.

The amount of reddening depends on the contraction rate
of the universe and hence on the equation
of state of the dominant form of matter.
As noticed in \cite{Wands,FB2,Wands2}, the reddening of the
spectrum has the right strength to turn a vacuum spectrum
on sub-Hubble scales into a scale-invariant spectrum
on super-Hubble scales if the universe is dominated by
cold matter. More specifically, on scales which exit the
Hubble radius in a matter-dominated phase, an initial
vacuum spectrum on sub-Hubble scales is converted into
a scale-invariant one on super-Hubble scales. This
observation was used to propose the ``matter
bounce" alternative to cosmological inflation for creating
a scale-invariant spectrum of cosmological fluctuations. Such
a matter bounce is naturally realized \cite{LWbounce} in the context of
the ``Lee-Wick" model \cite{LW} for scalar field matter,
which is a particular case of the more general quintom matter
bounce scenario \cite{quintombounce}. Modified
gravity theories such as the Biswas et al. \cite{Biswas1}
ghost-free higher derivative gravity theory or Ho\v{r}ava-Lifshitz
gravity \cite{Horava} on spatially non-flat spatial sections can also
lead to matter bounce scenarios, as studied in \cite{Biswas2}
and \cite{Horavabounce}, respectively \footnote{In the case of
the theory of \cite{Biswas1}, the analysis of \cite{Biswas2},
which demonstrates that string gas matter \cite{SGC}
does not prevent a cosmological bounce, can easily
be generalized to construct a matter bounce.}.

In this Note we will compute the change in the amplitude
and slope of the spectrum of cosmological perturbations
from one cycle to the next. We begin with a short review of
the relevant formalism. Then, we compute the change in the
spectrum of cosmological perturbations from one cycle
to the next.

\section{Fluctuations in a Cyclic Background Cosmology}

We postulate the existence of a cyclic background cosmology.
The turnaround between the expanding phase and the contracting
phase at large radius could be generated by a spatial curvature
term (in the absence of a cosmological constant), for the turnaround
between the contracting phase and the expanding phase new ultraviolet
(UV) physics which violates the weak energy condition is required. 
In the context of the Einstein field equations, such new UV
physics can be modeled by quintom matter \cite{quintom}. 
Asymptotically free higher derivative gravity actions such as
the ones proposed in \cite{nonsingular,Biswas1} can also lead
to a non-singular bounce. Finally, in a background with non-vanishing
spatial curvature, Ho\v{r}ava-Lifshitz gravity \cite{Horava} can also lead to
a bounce under the conditions on matter spelled out in \cite{Horavabounce}.

In Figure 1 we present a space-time sketch of a cyclic cosmology.
We choose the origin of the time coordinate (vertical axis) to coincide
with a bounce point. The equations simplify if we work in terms of
conformal time $\eta$ which is related to the physical time $t$ via
$dt \, = \, a(\eta) d \eta$, where $a(\eta)$ is the scale factor. During
the time interval between $- \eta_c$ and $\eta_c$, the new UV physics
which yields the non-singular bounce is dominant, for other times
the effective equations of motion for gravity are assumed to reduce
to those of Einstein gravity.

\begin{figure}[htbp]
\includegraphics[scale=0.3]{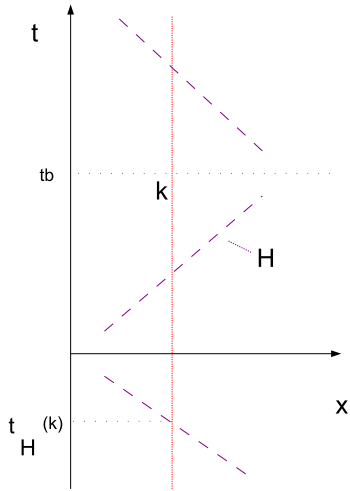}
\caption{Space-time sketch of the cyclic background cosmology. Time
is along the vertical axis, comoving length along the horizontal axis. The
majenta curve (labelled by $H$) denotes the Hubble radius $H^{-1}(t)$, 
the vertical line labelled by $k$ represents the wavelength of a perturbation
mode which exits the Hubble radius at the time $t_H(k)$. The origin
of the time axis is chosen to be the non-singular bounce point (around
the bounce the Hubble radius diverges to infinity, an effect not shown
on the graph. The time $tb$ is the turnaround from the expanding phase
to the contracting phase.}
\end{figure}

The horizontal axis in Figure 1 denotes co-moving distance $x$. The vertical
line corresponds to the wavelength of a cosmological fluctuation
mode. This mode crosses the Hubble radius (the dashed curve) at times
$\pm \eta_H(k)$. 

In the presence of cosmological perturbations without anisotropic stress,
the metric in longitudinal gauge takes the form
\be
ds^2 \, = \, a^2(\eta) \bigl[ (1 + 2 \Phi) d\eta^2 - (1 - 2 \Phi) d{\bf{x}}^2 \bigr] \, ,
\ee
where the function $\Phi(x, \eta)$ describes the fluctuations (see e.g. \cite{MFB}
for a review of the theory of cosmological perturbations). We are 
interested in computing the spectrum of $\zeta$, the function
describing the curvature fluctuations in co-moving coordinates.
$\zeta$ is given in terms of $\Phi$ via
\be
\zeta \, = \, \frac{2}{3} \bigl( {\cal{H}} \Phi^{'} + \Phi \bigr) \frac{1}{1 + w} + \Phi \, ,
\ee
$\cal{H}$ denoting the Hubble expansion rate in conformal time, a prime
indicating the derivative with respect of $\eta$, and $w = p / \rho$ being
the equation of state parameter of matter ($p$ and $\rho$ are pressure and
energy density, respectively).
The variable $\zeta$ in turn is closely related to the variable $v$ 
\cite{Sasaki,Mukhanov} in terms of which the action for
cosmological fluctuations has canonical kinetic term:
\be
\zeta \, = \, \frac{v}{z} \, .
\ee
Here, $z(\eta)$ is a function of the background which for constant
equation of state is proportional to the scale factor $a(\eta)$.

At linear order in perturbation theory, the equation of motion for the 
Fourier mode $v_k$ of $v$ is that of a harmonic oscillator with
a mass whose time-dependence is given by the background cosmology:
\be \label{EOM}
v_k^{''} + \bigl( k^2 - \frac{z^{''}}{z} \bigr) v_k \, = \, 0 \, .
\ee
This shows that, whereas on sub-Hubble scales $v_k$ is oscillating
with approximately constant amplitude, on length scales larger than 
the Hubble radius (where the $k^2$ term is negligible) the oscillations
of $v$ freeze out and the time-dependence of $v$ is given by
that of the background. One of the solutions of (\ref{EOM}) on
super-Hubble scales evolves as
$v \sim z \sim a$ and corresponds to a constant value of $\zeta$.
The second mode of $v$ corresponds to a decreasing mode
of $\zeta$ in an expanding universe. On the other hand,
in a contracting phase it corresponds to an increasing mode.

We see that the evolution of the curvature fluctuation $\zeta$ is
asymmetric between the contracting phase and the
expanding phase. The dominant mode is constant on
super-Hubble scales in the expanding phase whereas
it is increasing in the contracting phase. 
It is this asymmetry which is responsible
for the processing of the spectrum of fluctuations in a cyclic
background cosmology.

\section{Processing of the Spectrum of Cosmological Perturbations}

In this section, we will study the processing of the spectrum of
cosmological perturbations from one cycle to the next. We will
consider scales for which the background is expanding (or
contracting) as a power of time:
\be
a(t) \, \sim \, t^p \, .
\ee
In this case, the solutions of (\ref{EOM}) are given by
\be
v(\eta) \, \sim \, \eta^{\alpha}
\ee
with 
\be \label{alpha}
\alpha \, = \, \frac{1}{2} \pm \nu \,\,\,\, , \,\,\,\, \nu \, = \, \frac{1}{2} \frac{1 - 3p}{1 - p} \, .
\ee

We are interested in the range of values $1/3 < p < 1$ for which $\nu$ is a negative
number. In the contracting phase prior to the bounce at $\eta = 0$,
the dominant solution of  (\ref{EOM}) thus scales as 
\be \label{res}
v_k(\eta) \, \sim \, \eta^{1/2 + \nu} \, \sim \, \eta^{\frac{1- 2p}{1 - p}} \, .
\ee
There are two cases of special interest. First, in a matter-dominated universe $p = 2/3$
and hence the dominant mode of $v_k$ scales as
\be
v_k(\eta) \, \sim \, \eta^{-1} \, ,
\ee
whereas in a universe dominated by relativistic radiation $p = 1/2$ and
hence 
\be
v_k(\eta) \, \sim \, {\rm const} \, .
\ee
In fact, from (\ref{res}) we see that for matter with an equation of state
$w > 1/3$ the amplitude of $v$ is decreasing as the
bounce is approached, whereas for $w < 1/3$ the amplitude is increasing.
We will focus on the more physical second case.

Let us assume initial conditions for fluctuations on sub-Hubble scales
at some initial time $- \eta_i$ long before the bounce point:
\be
v_k(- \eta_i) \, = \, z(- \eta_i) A_i^{1/2} k^{-3/2} \bigl( \frac{k}{k_0} \bigr)^{(n_i - 1)/2} \, ,
\ee
where $A_i$ is the initial amplitude of the spectrum of $\zeta$, $n_i$ is the
initial slope, and $k_0$ is the pivot scale. Our aim is to calculate the
amplitude $A_f$ and slope $n_f$ after one cycle.

Making use of the fact that $v_k$ oscillates until the
time $- \eta_H(k)$ when the wavelength exits the Hubble radius
and subsequently increases in amplitude as given by (\ref{res}), we see that at
the time $- \eta_c$ immediately before the bounce, the time when the
weak energy violating effects which yield the non-singular bounce start
to dominate, the amplitude of $v_k$ is given by
\be
v_k(-\eta_c) \, = \, \bigl( \frac{- \eta_H(k)}{|\eta_c|} \bigr)^{(2p -1)/(1 - p)} v_k(\eta_i) \, .
\ee
Making use of the Hubble radius crossing condition 
\be \label{scaling}
\eta_H(k) \, \sim \, k^{-1} \, ,
\ee
we obtain the following power spectrum of $\zeta$ just before the bounce:
\bea \label{res1}
P_{\zeta}(k, -\eta_c) \, &\simeq& \, \bigl( \frac{z(- \eta_i)}{z(-\eta_c)} \bigr)^2 
\bigl( \eta_c k \bigr)^{-2(2p -1)/(1 -p)} A_i \bigl( \frac{k}{k_0} \bigr)^{n_i - 1} \nonumber \\
&=& \, A_f \bigl( \frac{k}{k_0} \bigr)^{n_f- 1} \, ,
\eea
with
\be \label{schange}
n_f \, = \, n_i - 2 \frac{2p -1}{1 - p}
\ee
and
\be \label{achange}
A_f \, = \, \bigl( \frac{z(- \eta_i)}{z(-\eta_c)} \bigr)^2 A_i \bigl( \eta_c k_0 \bigr)^{-2(2p -1)/(1 -p)} \, .
\ee
 
The next step of the analysis is to follow $v_k$ through the non-singular bounce from
time $-\eta_c$ to $\eta_c$. The first approach to do this would be to match the
values of the fluctuations at the two times using the Hwang-Vishniac \cite{HV} (see
also \cite{DM}) matching conditions. However, the applicability of this prescription
is questionable \cite{Durrer} since the background does not satisfy these matching
conditions. In the case of a non-singular bounce we can, however, follow the
evolution of the fluctuations through the bounce explicitly, assuming the validity
of the Einstein equations for the fluctuations. Since we are dealing with modes
which are in the far infrared (even at the bounce point) compared to the characteristic
scale of the bounce, this assumption is a safe one to make (as has been verified
explicitly in \cite{Biswas2} for the bounce model of \cite{Biswas1}). The lesson
that has been learned by following fluctuations through the bounce in several
models \cite{tsujikawa,omidbounce,tirthobounce,quintombounce} 
is that the spectrum of $v_k$
does not change on scales for which the wavelength is long compared to the
duration of the bounce. This is clearly satisfied for the case of interest to us.

In the expanding phase, the amplitude of $\zeta_k$ is constant on super-Hubble
scales. After the wavelength re-enters the Hubble radius, the amplitude
of $v_k$ does not change until the mode leaves the Hubble radius once more
in the contracting phase of the following cycle. To obtain the change in
the spectrum of cosmological perturbations, we need to evaluate the
spectrum of fluctuations at the time $\eta_i$, the mirror image of the
initial time. The spectrum at $\eta_i$ will be identical to the spectrum at
the time corresponding to $- \eta_i$ in the next cycle.

To compute this spectrum, note that the amplitude of $\zeta_k$ is decreasing
between the time $\eta_H(k)$ when the mode re-enters the Hubble radius
and the time $\eta_i$ since it is $v_k$ which has constant amplitude during
this time interval. Thus
\be
P_{\zeta}(k, \eta_i) \, = \, \bigl( \frac{z(\eta_H(k))}{z(\eta_c)} \bigr)^2
P_{\zeta}(k, \eta_c) \, .
\ee
Making use of the fact that $z$ scales as $\eta^{p/(1 - p)}$ and of the
relations (\ref{res1}) and (\ref{achange}) we find
\bea \label{res2}
P_{\zeta}(k, \eta_i) \, &=& \, \bigl( \frac{\eta_H(k)}{\eta_c} \bigr)^{2p/(1-p)} 
\bigl( \eta_c k_0 \bigr)^{-2 \frac{2p -1}{1 -p}} A_i \bigl( \frac{k}{k_0} \bigr)^{n_f - 1} \nonumber \\
&=& \, A_F \bigl( \frac{k}{k_0} \bigr)^{n_F - 1} \, ,
\eea
with
\be \label{schange2}
n_F \, = \, n_i - 2 \frac{3p -1}{1 - p}
\ee
and
\be \label{achange2}
A_F \, = \, A_i \bigl( \eta_c k_0 \bigr)^{-2(3p -1)/(1 -p)} \, .
\ee
 
The results
(\ref{schange2}) and (\ref{achange2}) give the change in the slope and in the
amplitude of the spectrum of cosmological perturbations from one cycle
to the next. Each cycle leads to a reddening of the spectrum and to an 
increase in its amplitude. The slope changes by
$-2(3p - 1)/(1 -p)$, The increase
in amplitude soon leads to a breakdown of the validity of the perturbative
analysis, with consequences for the multiverse discussed in \cite{piao}
 
As mentioned in the Introduction, if $p = 2/3$, then an initial vacuum
spectrum ($n = 3$) on sub-Hubble scales is transformed into a
scale-invariant spectrum after the bounce. If we want an initial
Poisson spectrum ($n = 4$) to be transformed into a scale-invariant spectrum
after one complete cycle in the second expanding phase, the
background needs to satisfy $p = 7/13$, i.e. an equation of state
close to that of radiation. This is a simple application of the
processing of the spectrum of cosmological perturbations which we
have discussed here.
 
\section{Conclusions and Discussion}

We have studied the evolution of the linear cosmological fluctuations
from one cycle to the next in a cosmology with a periodic background.
Due to the asymmetry in the evolution of fluctuations in the contracting
and expanding phases, there is a large increase in the amplitude
of the perturbations from one bounce to the next. In addition, there
is a characteristic reddening of the shape of the spectrum.

The immediate implication of our analysis is that the evolution of
the universe in a cosmology with a cyclic background is not
cyclic. 

The characteristic processing of the slope of the spectrum
of perturbations which we discuss in this note in principle opens
up new avenues of generating a scale-invariant spectrum of
perturbations during a specific cycle. For example, a matter-dominated
background with a spectrum with an initial steep blue index $n_s = 5$ 
will yield a scale-invariant spectrum after two bounces, in the same way
that an initial vacuum spectrum with $n_s = 3$ yields a scale-invariant
spectrum after one bounce \cite{FB2}. The
processing of fluctuations also opens up new ways of generating
a scale-invariant spectrum of curvature perturbations starting
from thermal inhomogeneities \cite{thermal}. Note that a characteristic
feature of such scenarios are large non-Gaussianities with a particular
shape, as worked out in \cite{matterbounceng}.

However, the increase in the amplitude of the fluctuations quickly leads to
a breakdown of the linear analysis. The nonlinear evolution will then
lead to entropy production and to the usual problems for cyclic
background cosmologies first discussed in \cite{Tolman}.

We should note that the cyclic version \cite{Cyclic} of the Ekpyrotic
universe scenario does not suffer from the problems discussed here
since the scale factor of ``our" universe in the Ekpyrotic model \cite{Ekp}
is not cyclic. The only cyclicity in the Ekpyrotic scenario is in the evolution
of the distance between the two boundary branes. The scale factor
of our universe is monotonically increasing. Thus, perturbations produced
at a fixed physical scale in one ``cycle" are redshifted by the time the
next ``cycle" arrives. 

\begin{acknowledgments} 
 
I wish to thank Y.-S. Piao for discussions which motivated this
study. I am grateful to
the Theory Division of the Institute of High Energy
Physics in Beijing and the CERN Theory Division for hospitality
and support during my sabbatical year.
This research has been supported by an NSERC Discovery Grant and 
by the Canada Research Chairs Program. 
 
\end{acknowledgments}


\begin{thebibliography}{99} 
 
 \bibitem{Novello}
M.~Novello and S.~E.~P.~Bergliaffa,
  ``Bouncing Cosmologies,''
  Phys.\ Rept.\  {\bf 463}, 127 (2008)
  [arXiv:0802.1634 [astro-ph]].
   
 \bibitem{Cyclic}
 P.~J.~Steinhardt and N.~Turok,
  ``Cosmic evolution in a cyclic universe,''
  Phys.\ Rev.\  D {\bf 65}, 126003 (2002)
  [arXiv:hep-th/0111098].
  
 \bibitem{Ekp}
 J.~Khoury, B.~A.~Ovrut, P.~J.~Steinhardt and N.~Turok,
  ``The ekpyrotic universe: Colliding branes and the origin of the hot big
  bang,''
  Phys.\ Rev.\  D {\bf 64}, 123522 (2001)
  [arXiv:hep-th/0103239].
  
 \bibitem{Tolman}
 R.~C.~Tolman,
  ``On the Problem of the Entropy of the Universe as a Whole,''
  Phys.\ Rev.\  {\bf 37}, 1639 (1931).

 \bibitem{Biswas3}
 T.~Biswas,
  ``The Hagedorn Soup and an Emergent Cyclic Universe,''
  arXiv:0801.1315 [hep-th].
  
 \bibitem{BMP}
 R.~H.~Brandenberger, T.~Prokopec and V.~F.~Mukhanov,
  ``The Entropy Of The Gravitational Field,''
  Phys.\ Rev.\  D {\bf 48}, 2443 (1993)
  [arXiv:gr-qc/9208009];\\
  R.~H.~Brandenberger, V.~F.~Mukhanov and T.~Prokopec,
  ``Entropy of a classical stochastic field and cosmological perturbations,''
  Phys.\ Rev.\ Lett.\  {\bf 69}, 3606 (1992)
  [arXiv:astro-ph/9206005].
  
 \bibitem{Wands}
 D.~Wands,
  ``Duality invariance of cosmological perturbation spectra,''
  Phys.\ Rev.\  D {\bf 60}, 023507 (1999)
  [arXiv:gr-qc/9809062].

\bibitem{FB2}
F.~Finelli and R.~Brandenberger,
  ``On the generation of a scale-invariant spectrum of adiabatic  fluctuations
  in cosmological models with a contracting phase,''
  Phys.\ Rev.\  D {\bf 65}, 103522 (2002)
  [arXiv:hep-th/0112249].
  
\bibitem{Wands2}
 L.~E.~Allen and D.~Wands,
  ``Cosmological perturbations through a simple bounce,''
  Phys.\ Rev.\  D {\bf 70}, 063515 (2004)
  [arXiv:astro-ph/0404441].
 
 \bibitem{LWbounce}
 Y.~F.~Cai, T.~t.~Qiu, R.~Brandenberger and X.~m.~Zhang,
  ``A Nonsingular Cosmology with a Scale-Invariant Spectrum of Cosmological
  Perturbations from Lee-Wick Theory,''
  arXiv:0810.4677 [hep-th].

 \bibitem{LW}
 B.~Grinstein, D.~O'Connell and M.~B.~Wise,
  ``The Lee-Wick standard model,''
  Phys.\ Rev.\  D {\bf 77}, 025012 (2008)
  [arXiv:0704.1845 [hep-ph]].
  
 \bibitem{quintombounce}
  Y.~F.~Cai, T.~Qiu, Y.~S.~Piao, M.~Li and X.~Zhang,
  ``Bouncing Universe with Quintom Matter,''
  JHEP {\bf 0710}, 071 (2007)
  [arXiv:0704.1090 [gr-qc]];\\
 Y.~F.~Cai, T.~Qiu, R.~Brandenberger, Y.~S.~Piao and X.~Zhang,
  ``On Perturbations of Quintom Bounce,''
  JCAP {\bf 0803}, 013 (2008)
  [arXiv:0711.2187 [hep-th]].
  
 \bibitem{Biswas1}
 T.~Biswas, A.~Mazumdar and W.~Siegel,
  ``Bouncing universes in string-inspired gravity,''
  JCAP {\bf 0603}, 009 (2006)
  [arXiv:hep-th/0508194].

 \bibitem{Horava}
 P.~Horava,
  ``Membranes at Quantum Criticality,''
  JHEP {\bf 0903}, 020 (2009)
  [arXiv:0812.4287 [hep-th]];\\
P.~Ho\v{r}ava,
  ``Quantum Gravity at a Lifshitz Point,''
  Phys. Rev. {\bf D 79}, 084008 (2009)
 [arXiv:0901.3775 [hep-th]].
  
 \bibitem{Biswas2}
  T.~Biswas, R.~Brandenberger, A.~Mazumdar and W.~Siegel,
  ``Non-perturbative gravity, Hagedorn bounce and CMB,''
  JCAP {\bf 0712}, 011 (2007)
  [arXiv:hep-th/0610274].

 \bibitem{Horavabounce}
 R.~Brandenberger,
  ``Matter Bounce in Horava-Lifshitz Cosmology,''
  arXiv:0904.2835 [hep-th].
  
 \bibitem{SGC}
 R.~H.~Brandenberger and C.~Vafa,
  ``Superstrings in the Early Universe,''
  Nucl.\ Phys.\  B {\bf 316}, 391 (1989);\\
  A.~Nayeri, R.~H.~Brandenberger and C.~Vafa,
  ``Producing a scale-invariant spectrum of perturbations in a Hagedorn  phase
  of string cosmology,''
  Phys.\ Rev.\ Lett.\  {\bf 97}, 021302 (2006)
  [arXiv:hep-th/0511140];\\
 R.~H.~Brandenberger,
  ``String Gas Cosmology,''
  arXiv:0808.0746 [hep-th].
  
  \bibitem{quintom}
 B.~Feng, X.~L.~Wang and X.~M.~Zhang,
  ``Dark Energy Constraints from the Cosmic Age and Supernova,''
  Phys.\ Lett.\  B {\bf 607}, 35 (2005)
  [arXiv:astro-ph/0404224];\\
B.~Feng, M.~Li, Y.~S.~Piao and X.~Zhang,
  ``Oscillating quintom and the recurrent universe,''
  Phys.\ Lett.\  B {\bf 634}, 101 (2006)
  [arXiv:astro-ph/0407432];\\
  Z.~K.~Guo, Y.~S.~Piao, X.~M.~Zhang and Y.~Z.~Zhang,
  ``Cosmological evolution of a quintom model of dark energy,''
  Phys.\ Lett.\  B {\bf 608}, 177 (2005)
  [arXiv:astro-ph/0410654].

 \bibitem{nonsingular}
  R.~H.~Brandenberger, V.~F.~Mukhanov and A.~Sornborger,
  ``A Cosmological theory without singularities,''
  Phys.\ Rev.\  D {\bf 48}, 1629 (1993)
  [arXiv:gr-qc/9303001];\\
  V.~F.~Mukhanov and R.~H.~Brandenberger,
  ``A Nonsingular universe,''
  Phys.\ Rev.\ Lett.\  {\bf 68}, 1969 (1992).

 \bibitem{MFB}
V.~F.~Mukhanov, H.~A.~Feldman and R.~H.~Brandenberger,
  ``Theory of cosmological perturbations. Part 1. Classical perturbations. Part
  2. Quantum theory of perturbations. Part 3. Extensions,''
  Phys.\ Rept.\  {\bf 215}, 203 (1992);\\
  R.~H.~Brandenberger,
  ``Lectures on the theory of cosmological perturbations,''
  Lect.\ Notes Phys.\  {\bf 646}, 127 (2004)
  [arXiv:hep-th/0306071].

 \bibitem{Sasaki}
  M.~Sasaki,
  ``Large Scale Quantum Fluctuations in the Inflationary Universe,''
  Prog.\ Theor.\ Phys.\  {\bf 76}, 1036 (1986).

\bibitem{Mukhanov}
V.~F.~Mukhanov,
  ``Quantum Theory of Gauge Invariant Cosmological Perturbations,''
  Sov.\ Phys.\ JETP {\bf 67}, 1297 (1988)
  [Zh.\ Eksp.\ Teor.\ Fiz.\  {\bf 94N7}, 1 (1988)].
 
 \bibitem{HV}
 J.~c.~Hwang and E.~T.~Vishniac,
  ``Gauge-invariant joining conditions for cosmological perturbations,''
  Astrophys.\ J.\  {\bf 382}, 363 (1991).
  
 \bibitem{DM}
 N.~Deruelle and V.~F.~Mukhanov,
  ``On matching conditions for cosmological perturbations,''
  Phys.\ Rev.\  D {\bf 52}, 5549 (1995)
  [arXiv:gr-qc/9503050].

 \bibitem{Durrer}
 R.~Durrer and F.~Vernizzi,
  ``Adiabatic perturbations in pre big bang models: Matching conditions and
  scale invariance,''
  Phys.\ Rev.\  D {\bf 66}, 083503 (2002)
  [arXiv:hep-ph/0203275].

 \bibitem{tsujikawa}
 S.~Tsujikawa, R.~Brandenberger and F.~Finelli,
  ``On the construction of nonsingular pre-big-bang and ekpyrotic cosmologies
  and the resulting density perturbations,''
  Phys.\ Rev.\  D {\bf 66}, 083513 (2002)
  [arXiv:hep-th/0207228].
  
 \bibitem{omidbounce}
 R.~Brandenberger, H.~Firouzjahi and O.~Saremi,
  ``Cosmological Perturbations on a Bouncing Brane,''
  JCAP {\bf 0711}, 028 (2007)
  [arXiv:0707.4181 [hep-th]].
  
 \bibitem{tirthobounce}
 S.~Alexander, T.~Biswas and R.~H.~Brandenberger,
  ``On the Transfer of Adiabatic Fluctuations through a Nonsingular
  Cosmological Bounce,''
  arXiv:0707.4679 [hep-th].
  
  \bibitem{piao}
  Y.~S.~Piao,
  ``Proliferation in Cycle,''
  arXiv:0901.2644 [gr-qc].
  
\bibitem{thermal}
  Y.~F.~Cai, W.~Xue, R.~Brandenberger and X.~Zhang,
  ``Thermal Fluctuations and Bouncing Cosmologies,''
  arXiv:0903.4938 [hep-th].
  
\bibitem{matterbounceng}
 Y.~F.~Cai, W.~Xue, R.~Brandenberger and X.~Zhang,
  ``Non-Gaussianity in a Matter Bounce,''
  arXiv:0903.0631 [astro-ph.CO].
  
\end{thebibliography}
\end{document}